
 \magnification=1200

 \vsize=23.5 truecm
 \hsize=16 truecm

\nopagenumbers\parindent0pt\parskip13pt
\headline={\ifodd\pageno\rhl \else \lhl\fi}
\def\rhl{\tenrm\hfill\ -- \folio --}
\def\lhl{\tenrm-- \folio --\hfill}
\overfullrule 0pt

\font\smallrm=cmr8

\def\lsim{\raise0.3ex\hbox{$<$\kern-0.75em\raise-1.1ex\hbox{$\sim$}}}
\def\gsim{\raise0.3ex\hbox{$>$\kern-0.75em\raise-1.1ex\hbox{$\sim$}}}

\def\gev{{\rm GeV}}
\baselineskip 16pt
\pageno=0
\line{\hfill Preprint HU-TFT-94-6}
\line{\hfill 6 April 1994}

\vskip 2.5cm
\centerline{MULTIPLICITIES FOR LHC NUCLEAR COLLISIONS}
\centerline{USING HERA STRUCTURE FUNCTIONS}
\vskip 1cm
\centerline{K.J. Eskola$^1$, K. Kajantie$^2$
and P.V. Ruuskanen$^3$}
\vskip 1.5cm
\centerline{Abstract}
\vskip 0.6cm
We compute in QCD perturbation theory the transverse energy carried by
gluons, quarks and antiquarks with $p_T\ge p_0\approx 2$ GeV in Pb+Pb
collisions at $\sqrt{s}=5500$ $A$GeV by using structure functions compatible
with the small-$x$ increase observed at HERA.
This gives a perturbative estimate for the energy and entropy density of the
bulk system at times $\tau\sim 0.1$ fm.  The predicted initial gluon entropy
density gives a lower limit of about 2200...3400 for the final charged
multiplicity.  Sources of further entropy increase are discussed.
\vfill
\hrule\medskip
{\smallrm\baselineskip 13pt
\line {1) Department of Physics, P.O.Box 9,
00014 University of Helsinki, Finland;\hfill}
\line {kjeskola@ fltxa.helsinki.fi.\hfill}

\line {2) Department of Theoretical Physics, P.O.Box 9,
00014 University of Helsinki, Finland;\hfill}
\line {kajantie@phcu.helsinki.fi.\hfill}

\line {3) Research Institute for Theoretical Physics, P.O.Box 9,
00014 University of Helsinki, Finland;\hfill}
\line {vruuskanen@phcu.helsinki.fi.\hfill}
}
\eject
\pageno =1
\null\vskip 0.6cm
The new ep-data from HERA [1-2] imply an increase
in the structure function $F_2^p$ at small $x$
relative to the behaviour implied by earlier data. Distribution
function analyses using the new data have been carried out [3-4].
These analyses, in particular, constrain the gluon distribution function,
which is not directly measured.

The purpose of this letter is to estimate, using these new gluon distribution
functions, the average multiplicity in a central Pb+Pb collision at
Large Hadron Collider (LHC)
energies, $\sqrt{s}=5500$ $A$GeV.  As the first stage, we compute the
transverse energy carried by minijets, gluons, quarks and antiquarks, with
$p_T\ge p_0\approx 2$ GeV, produced in these collisions [5].  This calculation
is as reliable as calculations in lowest order perturbative QCD at a scale 2
GeV are: the main uncertainties are higher order corrections, which we account
for with a $K$ factor, and shadowing, which we neglect.  The second stage is
more model dependent.  We assume that the produced gluons form a system in
space-time at $\tau_i=1/p_0=0.1$ fm and calculate its energy density from
$\epsilon(\tau_i)=\langle E_T^A(|y|<0.5) \rangle \times p_0/(\pi R_A^2)$ and
estimate from this the entropy density.  Assuming that the further expansion
of the system is adiabatic, we obtain from this a lower limit for the total
multiplicity: later dynamical phenomena can only increase the entropy.

The expected charged multiplicity is a crucial variable for the planning
of experiments [6] and the theoretical predictions lie roughly in the
range 1500-8000. Our result with the new gluon distributions [3-4] enchanced
at small $x$ is $dN_{ch}/dy>2200$ for $xg(x,)\sim x^{-0.3}$
and $>3400$ for $xg(x)\sim x^{-0.5}$; for the old
distributions [7], used in ref. [5], with $xg(x,4\,{\rm GeV}^2)\sim$ const, the
limit is only $dN_{ch}/dy>910$.
Note that these estimates assume a $K$ factor of 2 and do not
take into account shadowing.
A reliable error estimate would require the study of higher order corrections
[8] and gluon shadowing [9] for each structure function set separately.
These tasks will not be pursued here.
However, some insight on the uncertainty is provided
by the dependence of our results on different structure function sets
and on the parameter $p_0$.
In any case, the new HERA distribution functions have  a
very interesting impact on the LHC heavy ion experiment.

Proceeding to the computation, all details and the physical motivation are
given in [5]. Given the distribution functions $f_{i/p}$, the basic quantities
calculated are the integrated jet cross section
$$
\sigma_{\rm jet}(p_0)=\int_{p_0^2}^{s/4}dp_T^2dy_1dy_2 {1\over2}
{d\sigma_{\rm jet} \over dp_T^2dy_1dy_2}, \eqno(1)
$$
and the first $p_T$ moment
$$
\langle E_T\rangle \sigma_{\rm jet}(p_0)=\int dp_Tdy_1dy_2 {d\phi\over2\pi}
{1\over2}{d\sigma_{\rm jet} \over dp_Tdy_1dy_2}(\epsilon_1+\epsilon_2)p_T,
\eqno(2)
$$
where $\epsilon_1=\epsilon(p_T,y_1)=1$ if the parton falls inside the
region of our interest and =0 otherwise. Especially, we are going to
consider the central rapidity region $|y|\le 0.5$.
The differential cross section is given by
$$
{d\sigma_{\rm jet} \over dp_T^2dy_1dy_2} =
\sum_{{ijkl=}\atop{q,\bar q,g}}
x_1f_{i/p}(x_1,p_T^2) x_2f_{j/p}(x_2,p_T^2)
{d\hat\sigma\over d\hat t}^{ij\rightarrow kl},\eqno(3)
$$
and the distribution functions used are  D0', H and D-' by MRS [3-4].
The behaviour of the gluon distributions of these sets at the scale
$Q=p_T=2$ GeV is illustrated in Fig.~1.

The results with $p_0 = 2$ GeV are given in Table 1, and the $p_0$ dependence
is studied in Figs. 2 and 3. The numbers of Table 1 are on the pp level,
to convert them to numbers appropriate for central $AA$ collisions we
also need the
value of the overlap function $T_{AA}({\bf b})$. We shall use
$T_{\rm PbPb}(0)=32/$mb. The estimate for the gluonic
contribution to initial $E_T$ in one unit of $y$ near $y=0$ is then
$$
\eqalign{
\langle E_T^A(|y|<0.5)\rangle&=T_{\rm PbPb}(0)\,\sigma_{\rm jet}(p_0)\langle
E_T\rangle\cr
&=\,\,\,2700\, {\rm GeV}\qquad {\rm  DO1:}\, xg(x)\sim{\rm const},\cr
&=\,\,\,4540\, {\rm GeV}\qquad {\rm  D0':}\,\,\,\, xg(x)\sim{\rm const},\cr
&=\,\,\,8640\, {\rm GeV}\qquad {\rm H:}\,\,\,\,\,\,\,\, xg(x)\sim x^{-0.3},\cr
&=15300\, {\rm GeV}\qquad  {\rm D-':}\, xg(x) \sim x^{-0.5}. \cr}\eqno(4)
$$
The increase in $E_T$ associated with the small-$x$ increase is
significant. However, this $E_T$ is the final $E_T$ only if no further
interactions take place in the system. The largest decrease in $E_T$ is
obtained if the system thermalises.

To obtain a lower limit for the final multiplicity we shall first assume that
the system thermalises immediately after formation at $\tau_i=0.1$ fm and
expands adiabatically thereafter.  Since the entropy is conserved in the
comoving frame, we can estimate the final multiplicity from the entropy at the
initial time $\tau_i$.  Aiming at a lower limit we shall only consider the
dominant gluonic component, which, as seen from Table 1, is about 80\% of the
total.  Then pressure, energy density and entropy density are
$p=aT^4$, $\epsilon=3aT^4$ and $s=4aT^3$ with $a=16\pi^2/90=1.75$. Since
the longitudinal size of the comoving volume is $\Delta z=\tau \Delta y$ we
can estimate $\epsilon(\tau_i)=\langle E_T^A(|y|<0.5)\rangle p_0/\pi R_A^2$ and
obtain for the charged final multiplicity
$$
\eqalign{
{dN_{ch}\over dy}&\approx{2\over3}{1\over3.6}{dS\over dy}\approx
{2\over3}{1\over3.6}\pi R_A^2 4aT_i^3\tau_i\cr
&={2\over3}{4\over3.6}\biggl[{1\over27}\pi R_A^2a\tau_i
\langle E_T^A(|y|<0.5)\rangle^3
\biggr]^{1/4} \cr
&=914\qquad  \,\,\,{\rm DO1,}\cr
&=1350\qquad  {\rm D0',}\cr
&=2180\qquad {\rm H, }\cr
&=3360\qquad {\rm D-'.}\cr
 }\eqno(5)
$$

The assumption of fast initial thermalization seems reasonable when new HERA
structure functions are used. In this case the
gluonic subsystem is already initially close to
chemical equilibrium, in the sense that the gluon number density is
close to the thermal density obtained by assuming that all the initial energy
density, $\epsilon(\tau_i)$, is thermalised. In other words, the average
energy of the gluons is close to $\epsilon/n\approx2.7T$ as in thermal gas of
bosons. The average energy per gluon within $|y|<0.5$ is obtained from
Table 1 as $\sigma_{\rm jet}\langle E_T(|y|<0.5)\rangle/2\sigma_{\rm jet}$.
The results for different structure functions are
$$
\eqalign{
E/{\rm gluon}
&=4.1\,{\rm GeV}\qquad 2.7T_i=2.0\,\gev\quad{\rm DO1,}\cr
&=3.6\,\gev\qquad\qquad    \, =2.2\,\gev\quad {\rm D0',}\cr
&=3.2\,\gev\qquad\qquad    \, =2.6\,\gev\quad {\rm H,}\cr
&=2.9\,\gev\qquad\qquad    \, =3.0\,\gev\quad {\rm D-'}.\cr }
\eqno(6)
$$
For the distributions with $xg(x)\sim{\rm const}$ the thermalisation of
gluons requires a degradation of their average energy by collisions. With
the HERA structure functions the secondary interactions are only needed for
changing the directions of momenta to make the distribution uniform.

The lower limit of multiplicity was obtained by considering only the gluonic
component and assuming immediate complete thermalisation among gluons.
Then the work done
against longitudinal expansion is maximal, much of the $E_T$ is shifted towards
larger rapidities and $\epsilon$ decreases $\sim \tau^{-4/3}$.  Maximal
multiplicity is obtained if the viscosity is strong enough to balance the work
done by the pressure [10].  In this case all $E_T$ remains in the interval
$|y|<0.5$ and the multiplicity can be estimated by using $E_T$/particle
$\approx$ 0.5 GeV. In a realistic intermediate situation one has to consider,
starting from the initial conditions in Table 1, the gluon, the quark and the
antiquark subsystems and their interactions [10,11] which tend to
thermalize the
system, first the gluons, then possibly the quarks and antiquarks.
Kinetic and chemical equilibria are attained to various degrees. All this
is associated with an increase in entropy and multiplicity, but quantitative
estimates depend on the details of modeling.

In summary, we have considered the effects of the increase of gluon
distributions in the small-x region as indicated
by the new HERA data on the predictions of perturbative QCD -- at the
small scale of 2 GeV -- on the partonic subsystem produced in heavy ion
collisions at LHC energies. With due reservation concerning possible
shadowing effects [9], the HERA results give a solid reason for expecting
rather large multiplicities and thus increase the likelihood of
observing new gluon-quark plasma phenomena at LHC.

\vskip1cm
\line{\it Acknowledgements\hfill} KK thanks F. Eisele for asking the question
which lead to this investigation.

\vfill\eject

\moveright2.5cm\vbox{
\halign{
#\hfil&\hfil#\qquad&\hfil#\qquad&\hfil#\qquad&\hfil#\qquad\cr
Set&$\sigma_{\rm jet}$&$g$&$q$&$\bar q$\cr
\noalign{\vskip6pt \hrule width8.65truecm \vskip6pt}
DO1\quad&\quad 130&\quad101&\quad18.8&\quad10.0\cr
{}\quad&\quad 12.1&\quad10.0&\quad1.12&\quad1.00\cr
{}&{}&{}&{}&{}\cr
D0'\quad&\quad 233&\quad183&\quad31.9&\quad18.5\cr
{}\quad&\quad 23.6&\quad19.5&\quad2.13&\quad2.00\cr
{}&{}&{}&{}&{}\cr
H\quad&\quad 512&\quad401&\quad70.5&\quad40.5\cr
{}\quad&\quad 50.9&\quad42.4&\quad4.31&\quad4.15\cr
{}&{}&{}&{}&{}\cr
D-'\quad&\quad1039&\quad833&\quad137&\quad68.9\cr
{}\quad&\quad93.3&\quad81.5&\quad6.02&\quad5.83\cr
}}
\vskip1.5cm
\moveright2.5cm\vbox{
\halign{
#\hfil&\hfil#\qquad&\hfil#\qquad&\hfil#\qquad&\hfil#\qquad\cr
Set&$\sigma_{\rm jet}\langle E_T\rangle$&$g$&$q$&$\bar q$\cr
\noalign{\vskip6pt \hrule width9.0truecm \vskip6pt}
DO1\quad&\quad 994&\quad787&\quad136&\quad70.5\cr
{}\quad&\quad 100&\quad84.5&\quad8.29&\quad7.41\cr
{}&{}&{}&{}&{}\cr
D0'\quad&\quad 1570&\quad1241&\quad212&\quad120\cr
{}\quad&\quad 170&\quad142&\quad14.5&\quad13.6\cr
{}&{}&{}&{}&{}\cr
H\quad&\quad 3100&\quad2430&\quad425&\quad241\cr
{}\quad&\quad 322&\quad270&\quad26.7&\quad25.5\cr
{}&{}&{}&{}&{}\cr
D-'\quad&\quad5890&\quad4720&\quad782&\quad392\cr
{}\quad&\quad549&\quad479&\quad35.8&\quad34.5\cr
}}
\vskip0.5cm
Table 1. Values of $\sigma_{\rm jet}(p_0)$ and $\sigma_{\rm jet}(p_0)\langle
E_T\rangle$
in units of mb and mbGeV calculated for $p_0=2$ GeV and for three
different MRS distribution functions with $xg(x,4{\rm GeV}^2)\sim$ const
(D0'), $\sim x^{-0.3}$ (H) and $\sim x^{-0.5}$ (D-') [3-4]. The results with
Duke-Owens set 1 distributions (DO1) [7] are shown for comparison.
An overall factor $K=2$ is included. Both the total as well
as the contribution of gluons, quarks and antiquarks is given. For each
set the upper numbers are for all $y$, the lower for $|y|<0.5$.

\vfill\eject
\null\vskip 1cm
\centerline{\bf References}
\vskip 0.4truecm\parindent15pt
{\baselineskip 12pt
\item{[1]} H1 collaboration, I. Abt et al., DESY report DESY 93-117
(August 1993).
\item{[2]} Zeus collaboration, M. Derrick et al., DESY report
DESY 93-110 (August 1993).
\item{[3]} A.D. Martin, W.J. Stirling and R.G. Roberts, Phys. Lett.
B306 (1993) 145.
\item{[4]} A.D. Martin, W.J. Stirling and R.G. Roberts, RAL preprint
93-077 (1993).
\item{[5]} K.J. Eskola, K. Kajantie and J. Lindfors, Nucl.Phys.
B323 (1989) 37.
\item{[6]} A Letter of Intent for A Large Ion Collider Experiment (ALICE),
CERN/LHCC/93-16.
\item{[7]}  D.W. Duke and J.F. Owens, Phys. Rev. D30 (1984) 49.
\item{[8]} S.D. Ellis, Z. Kunszt and D.E. Soper, Phys. Rev. D40 (1990) 2121.
\item{[9]} K.J. Eskola, J. Qiu and X.-N. Wang, Phys. Rev. Lett. 72 (1994) 36,
K.J. Eskola, Nucl. Phys. B400 (1993) 240.
\item{[10]} K.J. Eskola and M. Gyulassy, Phys. Rev. C47 (1993) 2329.
\item{[11]} T.S. Bir\'o, E. van Doorn, B. M\"uller, M.H. Thoma and X.-N. Wang,
Phys. Rev. C48 (1993) 1275.
}
\parindent0pt
\null\vskip 1cm
\centerline{\bf Figure Captions}
\medskip
{\bf Fig.~1.} Gluon distribution functions MRS D0', D-, and H [3-4] at the
scale $Q=2$ GeV.
The old gluon distribution of Duke-Owens set 1 (DO1) [7] is shown for
comparison.
\smallskip
{\bf Fig.~2.} The integrated jet  cross section $\sigma_{\rm jet}(p_0)$
computed from eq.~(1) with the sets of distribution functions marked on the
figure and for
$\sqrt{s}=200$ and 5500 GeV.
\smallskip
{\bf Fig.~3.} The first $p_T$ moment $\sigma_{\rm jet}(p_0) \langle E_T
\rangle$
from eq.~(2).
The curves are as in Fig.~2.

\bye